\documentclass[10pt]{article}
\usepackage{amsmath,amsthm,amssymb}
\usepackage[cp1251]{inputenc}
\usepackage[russian]{babel}
\usepackage{graphicx}
\usepackage[large]{caption2}
\usepackage{indentfirst}
\usepackage{amsmath}
\usepackage{afterpage}

\usepackage[usenames]{color}

\usepackage{amssymb,amsmath}
\usepackage{amsfonts}
\usepackage{epsfig}
\usepackage{ulem}

\usepackage{cite}
\def\gee{ \, \lower 1mm\hbox{$\,{\buildrel > \over{\scriptstyle\scriptstyle\sim} }\displaystyle \,$}}
\def\lee{ \, \lower 1mm\hbox{$\,{\buildrel < \over{\scriptstyle\scriptstyle\sim} }\displaystyle \,$}}
\def\|{\partial}

\def\varkappa {{\scriptstyle\partial}\! e}

\let\b=\baselineskip
\begin{document}\large
\renewcommand{\captionlabeldelim}{.}
\headheight 1.50true cm \headsep  0.7true cm \righthyphenmin=2

\parindent=10.5mm

\marginparwidth=20true mm

{\normalsize

\noindent Voronin A., Isaeva I., Khoperskov A., Grebenuk S. Decision Support System for Urbanization of the Northern Part of the Volga-Akhtuba Floodplain (Russia) on the Basis of Interdisciplinary Computer Modeling // Communications in Computer and Information Science, 2017, vol. 754, p. 419-429

\noindent{https://doi.org/10.1007/978-3-319-65551-2\_30}

}

\ 

\noindent\_\_\_\_\_\_\_\_\_\_\_\_\_\_\_\_\_\_\_\_\_\_\_\_\_\_\_\_\_\_\_\_\_\_\_\_\_\_\_\_\_\_

\ 

{\bf Decision support system for urbanization of the northern part of the Volga-Akhtuba floodplain (Russia) on the basis of interdisciplinary computer modeling}

Alexander Voronin$^{1,2}$, Inessa Isaeva$^{1}$, 
Alexander Khoperskov$^{1,3}$ and Sergey Grebenjuk$^{1}$

$^{1}$ Volgograd State University, Volgograd 400062, Russia, $^{2}$voronin@volsu.ru, $^{3}$khoperskov@volsu.ru, 

\textbf{Abstract}

There is a computer decision support system (CDSS) for urbanization of the northern part of the Volga-Akhtuba floodplain. This system includes subsystems of cognitive and game-theoretic analysis, geoinformation and hydrodynamic simulations.
The paper presents the cognitive graph, two-level and three-level models of hierarchical games for the cases of uncontrolled and controlled development of the problem situation.
We described the quantitative analysis of the effects of different strategies for the spatial distribution of the urbanized territories.
For this reason we conducted the territory zoning according to the level of negative consequences of urbanization for various agents.
In addition, we found an analytical solution for games with the linear dependence of the average flooded area on the urbanized area.
We  numerically  computed  a  game equilibrium for dependences derived from the imitational geoinformation and hydrodynamic  modeling  of flooding.
As the result, we showed that the transition to the three-level management system and the implementation of an optimal urbanization strategy minimize its negative consequences.

\textbf{Keywords}: {cognitive analysis, game theory, geoinformatics, numerical modeling, decision support system}

\section{Introduction}
The Volga-Akhtuba Floodplain (VAF) is a unique natural landscape in the lower reaches of the Volga River and its normal existence is determined by its spring high water.
Volzhskaya HPP (VHPP) usually regulates the hydrological regime of the Volga in the interests of hydropower industry, and it has significantly been reducing the spring water volume compared to natural flooding since 11 large dams appeared on the river \cite{gor:bos}.
In recent years, the Volga-Akhtuba plain has been flooded rarely more than 40\,\% of its total area.
Factors of floodplain territory dehydration are the limitation of flood peaks due to the requirements of hydrological safety of expanding agricultural and urbanized territories and natural and anthropogenic degradations of numerous small channels in the interfluve area.

Active and uncontrolled urbanization can significantly accelerate the degradation of the floodplain nature.
So, an actual problem is creating the scientifically based decision supporting system for the VAF territory development.
 This system should ensure the sustainability of the ecosystem, and other interests of population, business entities and authorities.
Here we represent the structure of the Decision Support System (DSS) for development of the northern part of the Volga-Akhtuba floodplain and its application for the management of the urbanization process.
Subsystems of cognitive and game-theoretic analysis, geoinformation and hydrodynamic simulation modeling are components of our DSS.
These subsystems we can use separately or in different combinations to solve various control tasks.
Our previous works \cite{vor:vas:pis,vas:vor:svet} describe the use of this DSS in the management of the flood hydrological regime and hydrotechnical projects in the northern part of the VAF.
\section{Structure of the Decision Support System}
\subsection{The subsystem of cognitive analysis}
This subsystem includes a module for PEST + E-analysis, and modules for the construction and analysis of cognitive maps.
We identified the following main groups of actors in the system: the collective agent (land buyers in the interfluve), the management centers (municipal and federal level), and the operational and development priorities (hydrological safety, environmental and socioeconomic priorities).
The general long-term priority is the preservation of the VAF ecosystem, and the short-term priority is the preservation of the flooding area.

We identified the following main factors that influence the urbanization of~VAF:
\begin{enumerate}
  \item The volume of spring flood.
  \item The condition of small channels of VAF.
  \item	The average flooded area ($S_f(S)$).
  \item	The area of economic lands.
  \item	The limit of flooded area.
  \item	The indicator of floodplain ecosystem condition ($\Phi$).
  \item	The level of economic activity in VAF.
  \item	The indicator of life's quality in VAF.
  \item	The level of infrastructure development.
  \item	The number of inhabitants in VAF.
  \item	The value of the objective function of the collective agent ($f_A$).
  \item	The area of the urbanized territory of VAF ($S$).
  \item	The selling price of land ($p$).
  \item	The value of the objective function of the municipal authority ($f_M$).
  \item	The value of the objective function of the federal center ($f_F$).
  \item	The fine rate ($R$).
  \item The maximum area offered for sale.
\end{enumerate}

The cognitive graph of an uncontrolled process contains 15 vertices from the
1st to the 14th and 17th in Figure 1. In the figure, we graph each link with a solid
line and a corresponding symbol ``+'' or ``--'', making a decision with a pointed line,
and fast interaction with a thin line, slow interaction with a bold line.
As we see, the graph contain unstable cycles for the fast interactions (3 -- 17 -- 12 -- 5 -- 3) and stable cycles for the slow interactions (3 -- 6 -- 8 -- 11 -- 12 -- 10 -- 7 -- 4 -- 2 -- 3), (3 -- 6 -- 16 -- 14 -- 13 -- 11 -- 12 -- 10 -- 7 -- 4 -- 2 -- 3), (3 -- 6 -- 8 -- 11 -- 12 -- 5 -- 3).

\begin{figure}[th]
\centering\includegraphics[width=0.8\hsize]{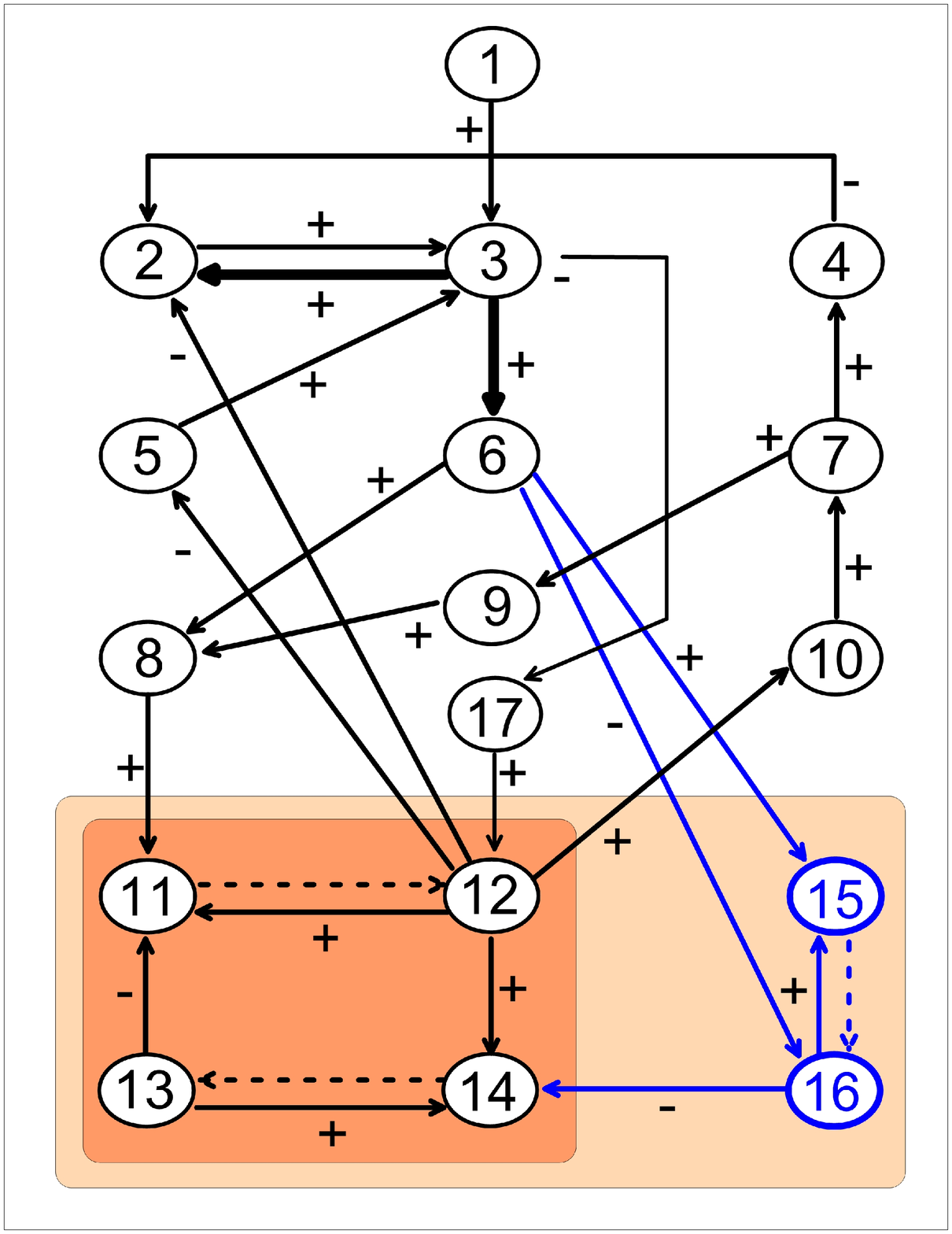}\vskip -0.\hsize
\caption{
A fragment of the cognitive graph of the VAF development for the cases of uncontrolled and controlled urbanization of the territory}
\label{voroninFig01}
\end{figure}

Our analysis of the graph shows that the slow stabilization of the urbanization process in the case of high effective demand leads to the depletion of the available territory and/or the decline the life quality in the VAF with the degradation of the ecosystem.
Vertexes 15, 16 and corresponding edges of the graph (see Figure\,\ref{voroninFig01}) describe elements of the mechanisms of environmental and economic management \cite{ahm:sim:wat,bur:kor:nov,nov:book,hag:klov,ougol,brun:hay:alt}. The goal of the management is to change ecological and socio-economic parameters of equilibrium.
\subsection{The subsystem of game-theoretic analysis}
The intent of the subsystem of game-theoretic analysis of the situation is a qualitative and quantitative assessment of the adequacy of the projected control mechanisms.
In Figure\,1 we marked a group of interacting vertices that are part of the game-theoretic model of the unmanaged process of buying and selling lands of the VAF with two participants.
The municipal authority is the seller, and the collective agent is the buyer.

For the two-level hierarchical game $\Gamma_1$ the mathematical formulation of the problem of finding the game equilibrium has the form:
\begin{equation}\label{eq-voronin-1}
\begin{array}{l}
  f_M=S p \rightarrow \max\limits_{p}\,, \\
  f_A = f_A^0(\Phi(S_f),S) - Sp \longrightarrow \max\limits_S \,, \\
  0\le S \le S_0\,, \ p\ge 0 \,,
\end{array}
\end{equation}
where $S_0$ is the total area of the Volga-Akhtuba floodplain, $f_A^{0}$ is the utility function of the collective agent.
 The solution of the problem (\ref{eq-voronin-1}) for functions $f_A^0 = a S_f S$, \ $\displaystyle S_f = - \frac{S_f^0}{S_0}\,S + S_f^0$ ($a\ge 0$ is a normalizing coefficient that takes demand into account, $S_f^{0} = S_f(0)$) has the form:
$$
    S = \frac{S_0}{4}\,,\quad S_f = \frac{3 S_f^0}{4}\,,
$$
\begin{equation}\label{eq-voronin-2}
 p = \frac{aS_f^0}{2} \,,
\end{equation}
$$
    f_M = \frac{a S_f^0 S_0}{8} \,, \quad f_A = \frac{a S_f^0 S_0}{16} \,.
$$
 For the northern part of the VAF we put $S_0=867$\,km$^2$, $S_f^0=220$\,km$^2$ and got the solution for the model (\ref{eq-voronin-1}) the solution $S=217$\,km$^2$, $S_f=165$\,km$^2$.

If we add to the model (\ref{eq-voronin-1}) the mechanisms of environmental and economic management implemented by the federal center, we can get a three-level hierarchical game:
\begin{equation}\label{eq-voronin-3}
\begin{array}{rcl}
f_F &=& f_F^0(\Phi) + R(\Phi) \longrightarrow \max\limits_R \,, \\
 f_M &=& Sp - R(\Phi) \longrightarrow \max\limits_p \,,
\\ \ \\
f_A &=& f_A^0(\Phi(S_f),S) - Sp  \longrightarrow \max\limits_S \,, \\
 0&\le& S\le S_0\,, \quad R(\Phi)\ge 0 \,,
\end{array}
\end{equation}
where $f_F^0$ is a function of the environmental utility of the federal center.

For functions
$$
 f_A^0 = aS_fS\,, \quad S_f = -\frac{S_f^0}{S_0} S + S_f^0 \,,
$$
$$
  f_F^0 = bS_f\,,\quad R=\lambda\left( S_f^0 - S_f \right)
$$
(where $b\ge 0$  is a normalizing factor that takes into account the environmental value of the VAF, $\lambda\ge 0$  is an optimized fine rate from the federal government) the solution of problem (\ref{eq-voronin-3}) is equilibrium in the Hermeyer's hierarchical game, and it has the form
\begin{equation}\label{eq-voronin-4}\displaystyle
\begin{array}{rcl}
S&=&\displaystyle\frac{S_0}{8}\left( 1-\varepsilon \right) \,,
\\ \ \\
 S_f &=& \displaystyle\frac{S_f^0}{8}(7+\varepsilon)\,,
\\ \ \\
 p&=& \displaystyle\frac{aS_f^0}{4}(3+\varepsilon) \,,
 \\ \ \\
 \lambda&=&\displaystyle\frac{aS_0}{2}(1+\varepsilon)\,,
\\ \ \\
 f_F&=&\displaystyle\frac{aS_f^0 S_0}{16}(1+14\varepsilon+\varepsilon^2)\,,
\\ \ \\
  f_M &=& \displaystyle\frac{a S_f^0 S_0}{32}(1-\varepsilon)^2\,,
\\ \ \\
 f_A&=&\displaystyle\frac{aS_f^0S_0}{64}(1-\varepsilon)^2\,,
 \\ \ \\
  \varepsilon &=& \displaystyle\frac{b}{aS_0} \quad (0\le \varepsilon \le 1) \,.
  \
\end{array}
\end{equation}
We adopted $a=5\cdot 10^{5}$\,RUB/km$^4$ for numerical estimates.

In the case of $\varepsilon \ge 1$  the solution $S=0$ is valid.
Comparison of (\ref{eq-voronin-2}) and (\ref{eq-voronin-4}) shows that the fines mechanism reduces twice the equilibrium value of the urbanized territory area and also it causes the growth of the average area of flooding even in the case of $\varepsilon=0$, if the penalty rate is determined by the price of land and  $\displaystyle\lambda=\frac{2pS_0}{3S_f^0}$.
In addition, the received amounts of fines can be used to finance environmental projects.
The municipal center can vary the strategies of offering land properties and determine the actual form of $S_f(S)$  according to the conditions of flooding and the ecological value of these land properties.
This function is convex if we use an aggressive urbanization strategy based on the order of the land sale according to their ecological value.
In the opposite case, we have a concave function  $S_f(S)$.

The exact form of the function $S_f(S)$ is determined by the features of the topography and the flooding regime of the VAF.
To simulate the dependences of $S_f(S)$ in the problems (\ref{eq-voronin-1}) and (\ref{eq-voronin-3}), we used here the subsystems of geoinformation and hydrodynamic modeling.
\subsection{The subsystem of geoinformation modeling}
The basis of our digital elevation model (DEM) for VAF in the form of heights matrix $b(x_i, y_j)$ are SRTM DEM and ASTER GDEM satellite data with a resolution of up to 20\,m in the earth's plane and up to 0.5\,m vertically.
To create the DEM we used the universal geoinformation system ``Panorama'' and special spatial data software \cite{vor:vas16}.
We perform an annual update of the DEM with new open satellite imagery of the US geological service (Landsat~8 Satellite), and with our own GPS measurements of flooding boundaries.

\begin{figure}[th]
\centering\includegraphics[width=1.0\hsize]{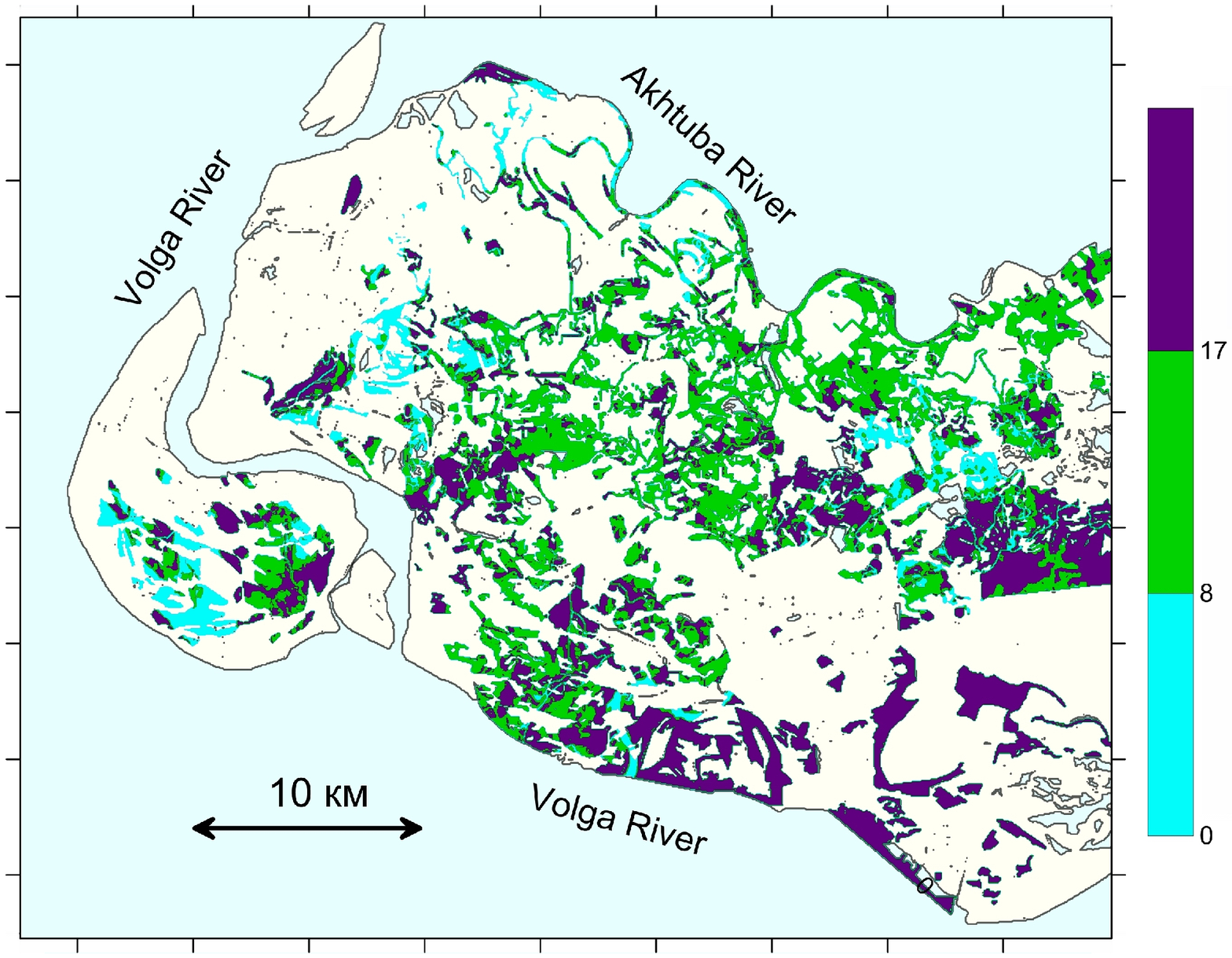}\vskip 0.05\hsize
\caption{
Map of the territory of the VAF. Different colors show areas of potential urbanization. Numbers show the non-flooding frequency of each map's point for the last 18 years}
\label{voroninFig02}
\end{figure}

The current VAF vector map includes a layer of the hydraulic system with 1542 riverbeds, a layer of infrastructure with 118 settlements, and a relief layer with more than 15,000 relief objects.
In addition to DEM, this subsystem includes a cadastral digital map and a cadastral database of VAF.
Figure~2 shows a digital map of VAF, where the areas of potential urbanization are highlighted with different colors.
\section{The subsystem of hydrodynamic modeling of floods}
We used a numerical model of the shallow water dynamics and we took into account all the main factors of the territory flooding \cite{khrap-etal13}:

\noindent --- mode of water supply through the Volga Hydroelectric Power Station (hydrograph);

\noindent --- surface and underground water sources;

\noindent --- terrain relief including anthropogenic development of the territory;

\noindent --- relief of the bottom of reservoirs;

\noindent --- properties of the underlying surface (bottom friction, infiltration);

\noindent --- internal viscous friction;

\noindent --- wind effect;

\noindent --- rotation of the Earth;

\noindent --- evaporation.

\begin{figure}[th]
\centering\includegraphics[width=1.0\hsize]{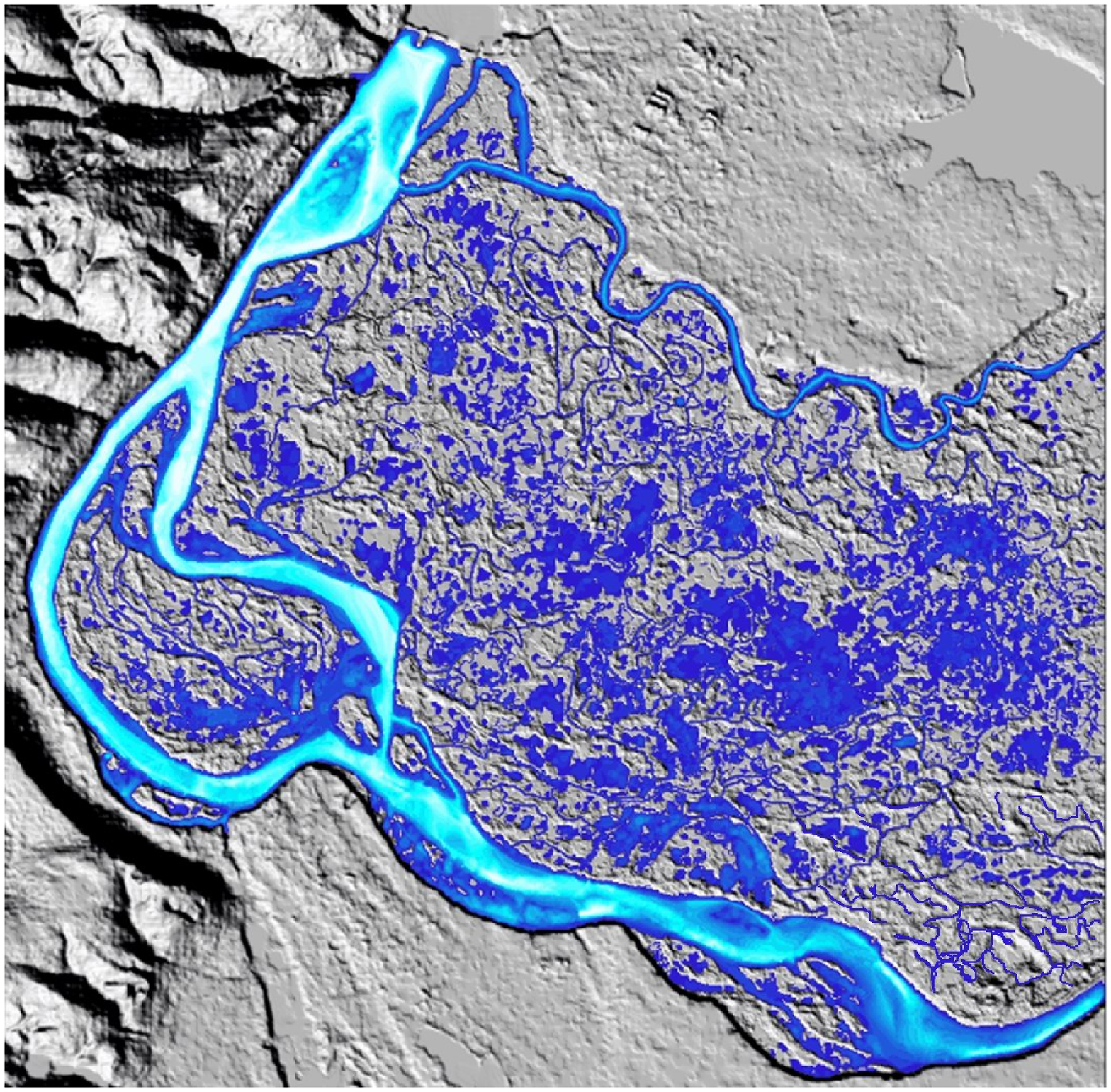}\vskip 0.05\hsize
\caption{
An example of the spring flood simulation in the northern part of the floodplain}
\label{voroninFig03}
\end{figure}

The result of numerical simulations of shallow water is a set of maps of water depth distribution at different times (Figure~3).
To assess the adequacy of the model we compared the results of numerical simulations with the water level observation data at four gauging stations in the northern part of the VAF and we checked the calculated water surface area in 2012--2016 against the Landsat-7 satellite data.
We implemented parallel OpenMP, CUDA, OpenMP-CUDA versions of the calculation module for the combined Smoothed Particle Hydrodynamics -- Total Variation Diminishing (CSPH-TVD) method, which significantly reduced the required computing resources \cite{dyak:khoper}.

\section{Simulation modeling of the urbanization process}
The subsystem of urbanization imitation modeling makes it possible to analyze the ecological consequences of decisions about the flooding of certain zones.
The information basis for analysis is a set of the digital maps of flooding with hydrographs from 1990 to 2016 and a digital cadastral map of the northern part of the VAF.
As the result of this module functioning we got the parameters determination for safe and dangerous hydrographs, and the maps of maximum safe flooding of a given part of the VAF.
We calculated the frequency of the territory flooding, as well as the predicted value of the average flooding area with guarantee no flooding of this territory.

\begin{figure}[th]
\centering\includegraphics[width=0.7\hsize]{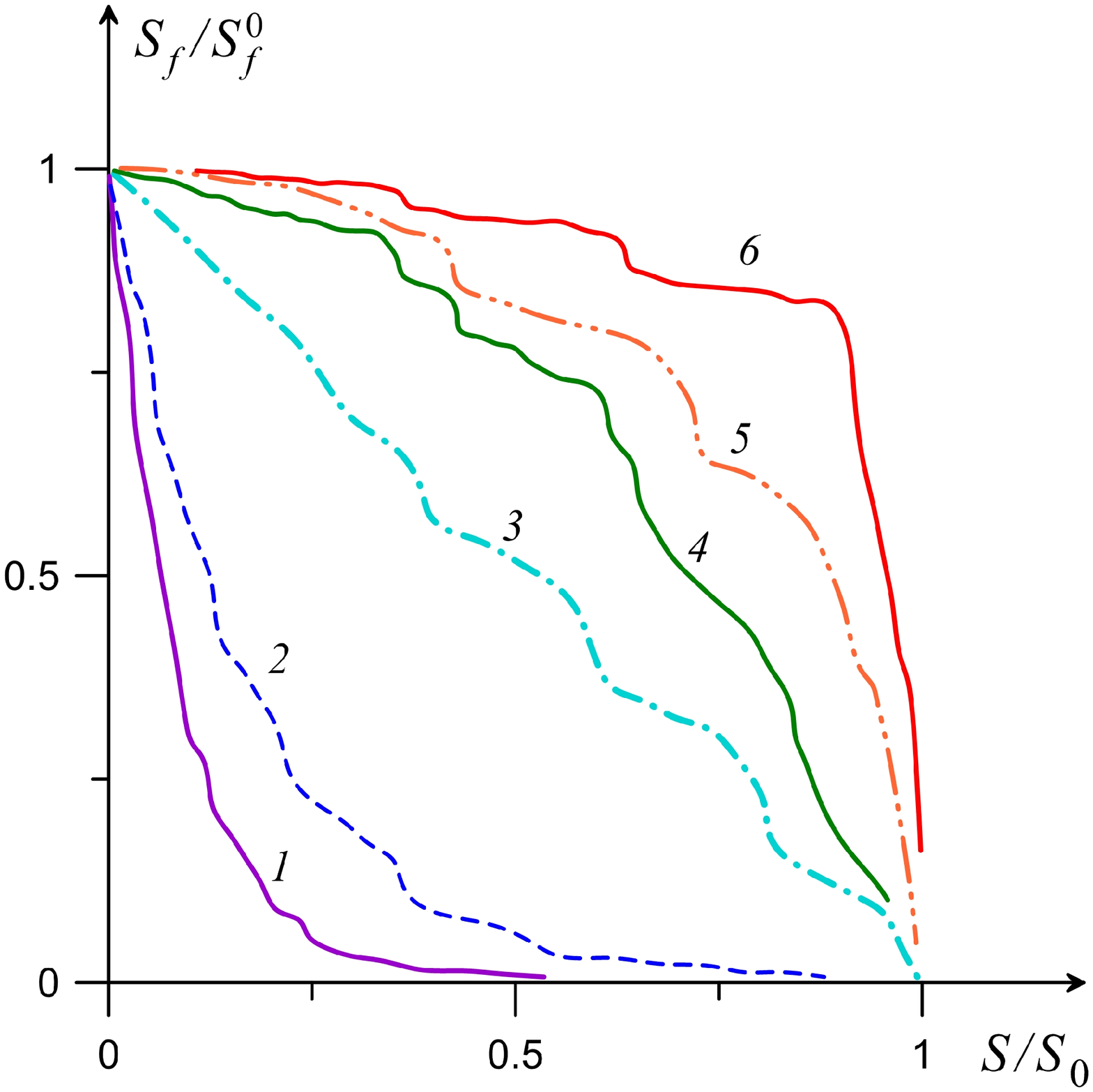}\vskip -0.\hsize
\caption{
Simulation dependencies for different urbanization strategies. Curves 1--6 are ordered according to the degree of ecological efficiency}
\label{voroninFig04}
\end{figure}

Let us discuss the results of the consequences analysis for the various urbanization strategies obtained in the interdisciplinary simulation (Figure~4).
The joint analysis of the cadastral and flood maps of the VAF allowed to estimate the frequency of flooding for various sites of potential urbanization.
To analyze the impact of urbanization strategies in the VAF on its flooding regime, we built the dependencies $S_f(S)$  for various strategies of lands marketing assignment. The strategies were ordered by the environmental friendliness.

Unsold territory was divided into three zones according to the frequency of flooding: 1~--- frequent event, 2~--- medium event, 3~--- rare event.
We numbered the grid nodes of the DEM within each zone randomly in accordance with the information uncertainty.
In Table~1 we indicated the ordering of zones in our strategies.
For example, in strategy~2, randomly selected areas from zone 1 are subsequently offered for sale. Then we repeat this procedure for zone~3 and then for zone~2.

\begin{table}
\caption{The ordering of the three zones in urbanization strategies}
\begin{center}
\begin{tabular}{|l|c|c|c|c|c|c|}
  \hline
  $\quad$ Strategy number $\quad$ & 1 & 2 & 3 & 4 & 5 & 6 \\
  \hline
  $\quad$ Zones order $\quad$ & \ 1 2 3 \ & \ 1 3 2 \ & \ 2 1 3 \ & \ 2 3 1 \ & \ 3 1 2 \ & \ 3 2 1 \ \\
  \hline
\end{tabular}
\end{center}
\end{table}

\begin{figure}[th]
\centering\includegraphics[width=0.7\hsize]{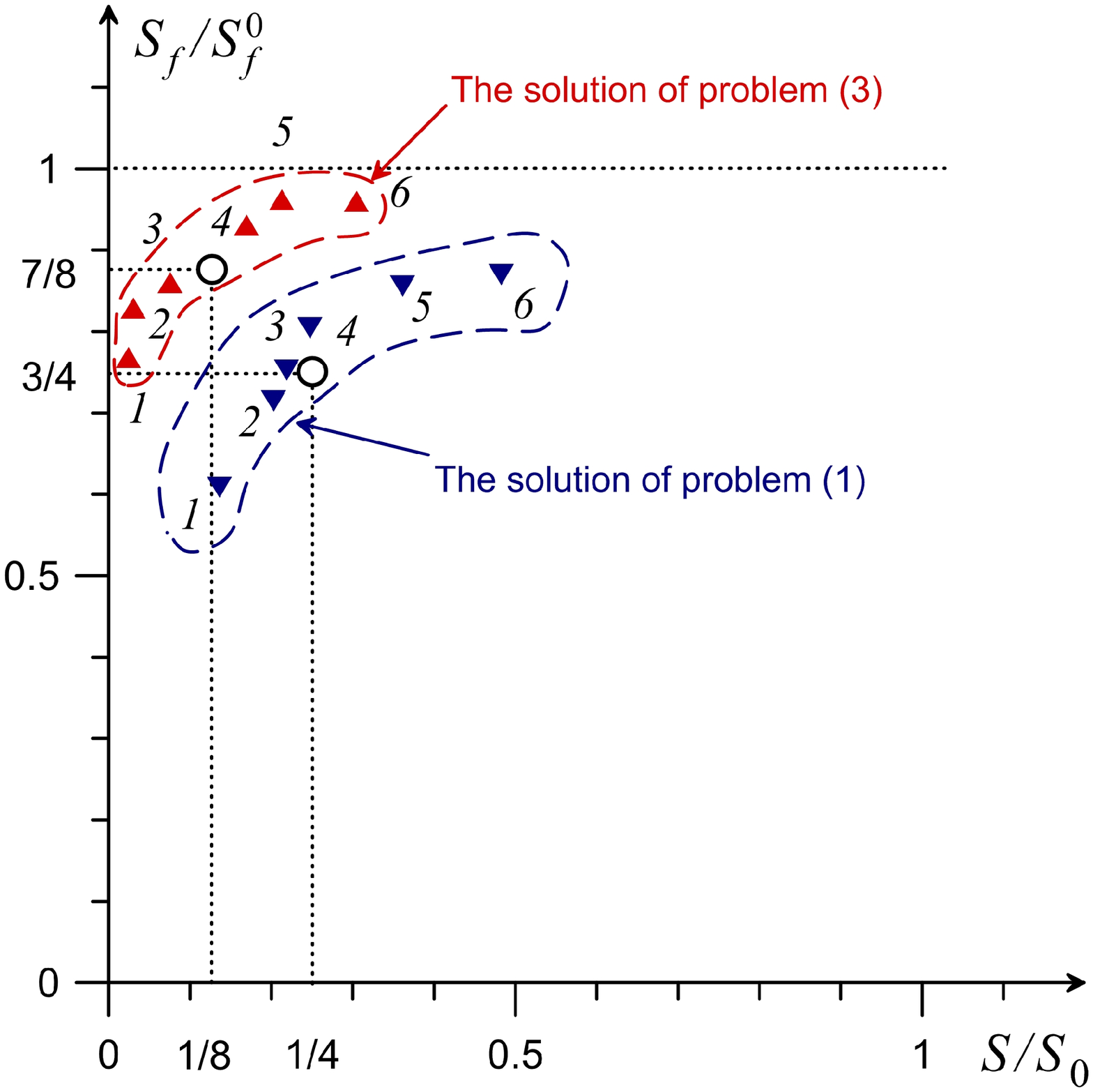}\vskip -0.\hsize
\caption{
The solutions of problems (\ref{eq-voronin-1}) and (\ref{eq-voronin-2}3) for the simulation dependencies $S_f(S)$  shown in Figure~4 for $\varepsilon=0$}
\label{voroninFig05}
\end{figure}

We calculated the average value of the flooding area using a set of flood maps for the years 1990\,--\,2016, but we excluded the maps where selected urbanized zone is flooded.
Figure~4 shows the curves of $S_f(S)$ for the strategies indicated in Table~1.
Then, for each of the dependencies $S_f(S)$  (see Figure~4) we found numerical solutions of problems (\ref{eq-voronin-1}) and (\ref{eq-voronin-3}) for $\varepsilon=0$.
Because of the difficulty in estimating the value of epsilon in numerical experiments we used lower or guaranteed estimates of the functions (\ref{eq-voronin-4}) corresponding to the value of $\varepsilon = 0$. With the growth of $\varepsilon$, the solution of problem (\ref{eq-voronin-3}) locates to the point (0;~1).
Figure~\ref{voroninFig05} shows the points that correspond to these solutions.
Circles indicate the positions of solutions (\ref{eq-voronin-2}) and (\ref{eq-voronin-4}).
Figure~5 demonstrates that the decrease of the equilibrium value $S$ in problem (\ref{eq-voronin-3}) stabilizes the unstable cycle in Figure~\ref{voroninFig01} for large values of the ecological criterion in comparison with problem (\ref{eq-voronin-1}). In addition, the use of environmental urbanization strategies leads to a significant increase in its effectiveness, both in terms of environmental and socio-economic criterion.

\section{Conclusion}
The presented simulation results show that the multidisciplinary Computer decision support system on the development of the territory of the Volga-Akhtuba floodplain allows us to evaluate the effectiveness of alternatives for the development of this territory.
If we add  hydrotechnical and nature restoration projects control subsystems, this will allow us to analyze the effectiveness of an integrated decision support system for the development of the floodplain territory.

\hfill \break
\noindent{\bf Acknowledgments.}
 AK is thankful to the Ministry of Education and Science of the Russian Federation (project No.\,2.852.2017/4.6).
 AV thanks the RFBR grant and Volgograd Region Administration (No.\,16-48-340147).

%
%

\end{document}